\documentclass[journal]{IEEEtran}
\usepackage{cite}
\usepackage{graphicx}
\usepackage[dvipsnames]{xcolor}
\usepackage{soul}
\usepackage{amsmath}
\usepackage{amssymb}
\usepackage{blindtext}
\usepackage{url}
\usepackage{xr-hyper}
\usepackage[colorlinks=true,allcolors=black]{hyperref}
\usepackage{array}

\DeclareUnicodeCharacter{2009}{} 
\DeclareUnicodeCharacter{2212}{} 

\makeatletter
\newcommand*{\addFileDependency}[1]{
 \typeout{(#1)}
 \@addtofilelist{#1}
 \IfFileExists{#1}{}{\typeout{No file #1.}}
}
\makeatother

\newcommand*{\myexternaldocument}[1]{%
 \externaldocument{#1}%
 \addFileDependency{#1.tex}%
 \addFileDependency{#1.aux}%
}
\myexternaldocument{si}

\begin{document}

\title{Order-of-Magnitude SNR Improvement for High-Field EPR Spectrometers via 3D-Printed Quasioptical Sample Holders}

\author{Anton\'{i}n~Sojka,
        Brad~D.~Price,
        Mark~S.~Sherwin

\thanks{This work of supported by the National Science Foundation (NSF) through the NSF research grants DMR 2117994 as well as MCB 2025860.

The authors are affiliated with the University of California, Santa Barbara, Santa Barbara, CA 93106, USA (e-mail: antonin\_sojka@ucsb.edu; sherwin@ucsb.edu).}}

\maketitle

\begin{abstract}  
In this paper, we present a rapidly-prototyped, cost-efficient, 3D-printed quasioptical sample holder for improving the signal-to-noise ratio (SNR) in modern, resonator-free, high-field electron paramagnetic resonance (EPR) spectrometers. Such spectrometers typically operate in induction mode: the detected EPR (``cross-polar") signal is polarized orthogonal to the incident (``co-polar") radiation. The sample holder improves SNR in three modes: continuous wave, pulsed, and rapid-scan. An adjustable sample positioner allows for optimizing sample position to maximize the 240\,GHz magnetic field $B_1$, and a rooftop mirror allows for small rotations of the cross-polar signal to maximize the signal and minimize the co-polar background.  When optimized, the co-polar isolation (the ratio of incident to detected co-polar signal) was  $\sim 50$\,dB, an improvement of $>20$\,dB. This large isolation is especially beneficial for maximizing the SNR of rapid-scan EPR, but also improves the SNR in pulsed and cwEPR experiments. Through minimal modification, the sample holder may be incorporated into a variety of homebuilt, induction-mode hfEPR spectrometers in order to significantly improve the SNR (approx. 6$\times$), and thereby reduce the acquisition time (by more than a factor of 30).
\end{abstract}


\IEEEpeerreviewmaketitle
\section{Introduction}

\IEEEPARstart{E}{lectron} paramagnetic resonance (EPR) is a spectroscopic technique based on the Zeeman effect. An unpaired electron spin in presence of external magnet field has two eigenstates, ``spin-up" and ``spin-down", and irradiating a spin-up electron with a photon of energy equal to the energy difference between the two states leads to photon absorption \cite{Weil2006}. EPR spectrometers make use of the Zeeman effect to probe the local environments of electron spins and can be found in a large fraction of chemistry and materials science laboratories, often to study radical species in a variety of environments (biological, solution, powder, crystal, etc.) or to characterize electronic materials \cite{freed_electron_1972, owens_paramagnetic_1985, farle_ferromagnetic_1987, che_chapter_1990, freed_new_2000,  gallez_assessment_2004, napolitano_isotropic_2008, davies_detection_2016, naveed_recent_2018, bonke_situ_2021}. Nowadays, EPR is also used in the development of novel materials such as single-molecule magnets \cite{hill_detailed_2002, schnegg_frequency_2009, datta_comparative_2009, ghosh_multi-frequency_2012, fataftah_trigonal_2020}, single-ion magnets \cite{sun_two-dimensional_2016, realista_mn_2016, pedersen_toward_2016, jenkins_coherent_2017, handzlik_identical_2020}, and quantum bits \cite{coronado_molecular_2020}; and thus is increasing in popularity. 

EPR spectrometers most often operate between $10$ and $100$\,GHz (with magnetic fields between 0.35 and 3.5\,T for $g\approx2$), with samples in resonant cavities to enhance signal-to-noise ratio (SNR). High-frequency (frequencies above 100\,GHz, fields above 3.5\,T) EPR (hfEPR), however, is advantageous for increased resolution, and is especially informative for systems that have gaps in their excitation spectra at zero magnetic field, including molecules with $S>1/2$ and large zero-field splittings \cite{earle_250-ghz_1993, lebedev_very-high-field_1994, andersson_examples_2003, fataftah_trigonal_2020}, collective spin excitations in antiferromagnets \cite{foner_high-field_1963, li_spin_2020}, as well as frustrated spin systems \cite{oseroff_magnetic_1982, martinho_magnetic_2001}. HfEPR spectrometers are also invaluable for measuring the spin dynamics of radicals used as polarizing agents in dynamic nuclear polarization (DNP)-enhanced NMR spectrometers, which operate above 7\,T \cite{darmstrong_200_2010, walker_temperature_2013}.


Motivated by the tremendous scientific opportunities, the development of hfEPR spectrometers is an active field of research \cite{eaton_high-field_1999, van_tol_quasioptical_2005, takahashi_pulsed_2012, neugebauer_ultra-broadband_2018, sojka_high-frequency_2020}. The well-known challenges in generating powerful electromagnetic radiation in the $100-1000$\,GHz band has resulted in delayed development of hfEPR as compared to low field EPR and NMR. Further, higher frequencies bring additional challenges: resonant cavities get smaller, and above 200\,GHz, most systems do not use a cavity at all, which forfeits the large signal-to-noise improvement they often provide \cite{froncisz_loop-gap_1982, walsh_enhanced_1986, tschaggelar_high-bandwidth_2017}. Importantly, hfEPR spectrometers typically operate in induction mode, meaning only polarization orthogonal to the excitation polarization is detected. While typically capable of isolating the excitation polarization by about 30\,dB, induction mode architectures still impart a significant background to EPR experiments, as the excitation power is typically much larger than the orthogonal induction mode signal; as a result, improving induction-mode isolation for high field-and-frequency experiments is a potential avenue for improving hfEPR resolution and is a field of interest for many researchers \cite{cruickshank_kilowatt_2009,reijerse_high-frequency_2009, stepanov_high-frequency_2015, sojka_high-frequency_2021, schubert_terahertz_2022, abhyankar_recent_2022}. 

\begin{figure}[htbp]
    \centering
    \includegraphics[width=0.8\linewidth]{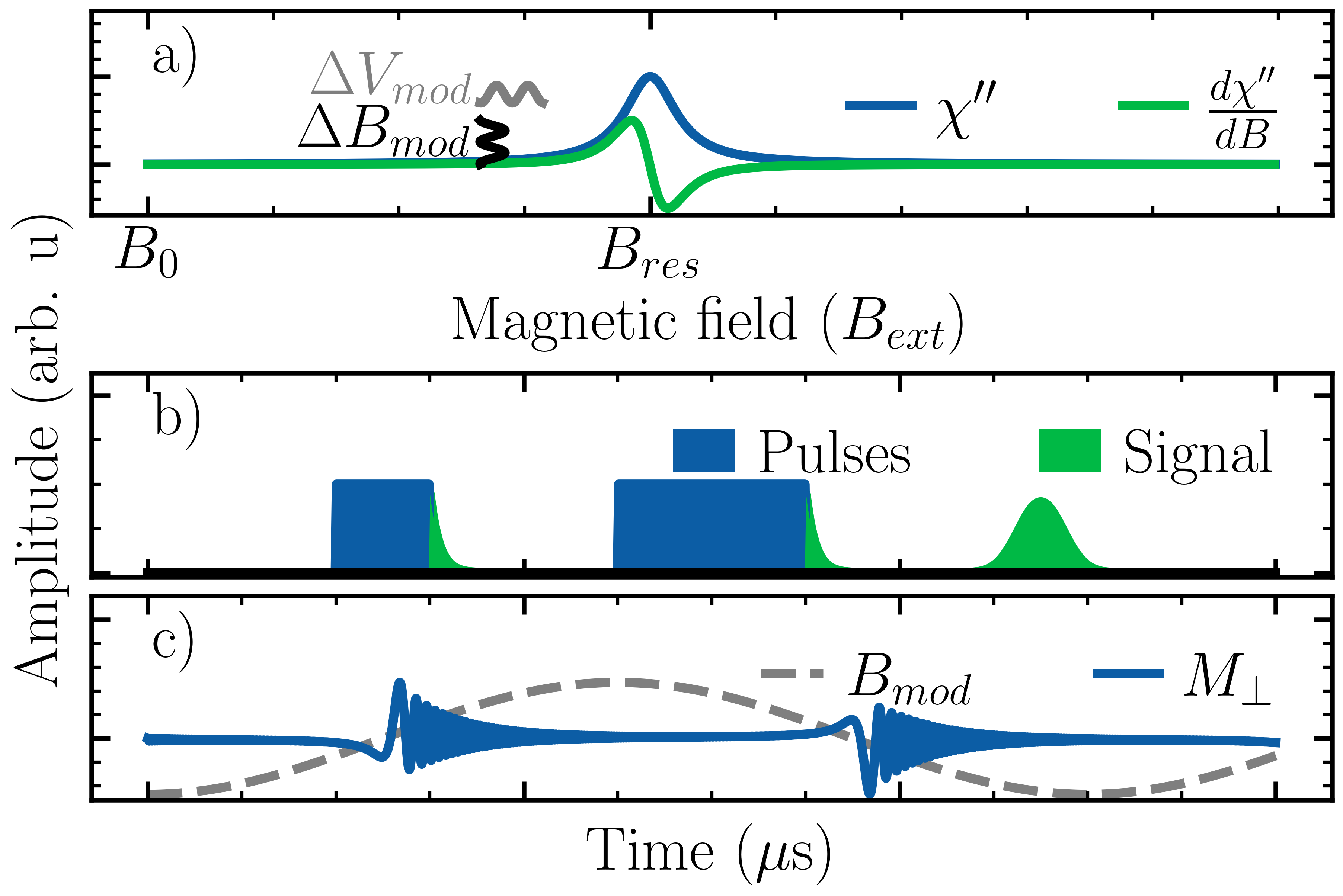}
    \caption{Simulated signals for continuous wave, pulsed, and rapid-scan EPR. (a) In cwEPR, the external magnetic field is modulated ($B_{mod}$) by a coil and the EPR signal is detected by a lock-in amplifier (resulting modulated signal is $V_{mod}$). The resulting EPR signal (green) is the derivative of the imaginary part of the complex magnetic susceptibility ($\chi''$, blue). (b) Pulsed EPR uses short driving pulses (blue) to excite the spins out of equilibrium and probe their dynamics; in the Hahn echo experiment shown here, the first pulse rotates the net magnetization of the spins, and a second pulse refocuses their precession in the rotating frame, resulting in a spin echo (green). (c) Rapid-scan EPR utilizes fast, large-amplitude modulation (black dashed line) through the resonance and direct-detection of the induction mode signal to record the response of the spins (blue and green solid lines). Fast repetition and continuous signal averaging can allow for significantly better SNR after shorter acquisition time than conventional cwEPR \cite{eaton_rapid-scan_2014}.}
    \label{fig:epr-modes}
\end{figure}

Most EPR systems operate in one (or more) of three modes: continuous wave (cw), where the resonance is swept through slowly, either by frequency or field; pulsed, with a fixed field and frequency, to obtain electron spin-lattice ($T_1$) or spin-spin ($T_2$) relaxation times; or rapid-scan, where a rapid sweep of field or frequency through the resonance distorts the EPR lineshape and can provide the slow-scan lineshape as well as $T_2$ relaxation time through post-processing (see Fig. \ref{fig:epr-modes}).

In this paper, we report the frequency-independent high-field sample holder which enables an important improvement in SNR for high-frequency EPR spectrometers that is easy to design, prototype, and implement.  The Quasioptical Sample Holder (QSH) presented here, designed to operate in all three modes (cw, pulsed, and rapid-scan), was fitted to attach to the EPR probe used in EPR spectrometer at the Institute for Terahertz Science and Technology (ITST). Minor adjustments to the 3D-printed model (.stp files included in S.I.) would allow for simple incorporation into other probe designs. The QSH was entirely 3D-printed for rapid prototyping, except for the 1/2" diameter parabolic mirror ($f=1"$), which was purchased from Thorlabs, Inc. \cite{noauthor_thorlabs_nodate}, and the rooftop mirror, which was end milled from a 7\,mm diameter aluminum rod (see S.I. Fig.
\ref{fig:roof-mirror}).

\section{Experimental Details}\label{sec:details}
\begin{figure}[htbp]
\begin{center}
\includegraphics[width=0.8\linewidth]{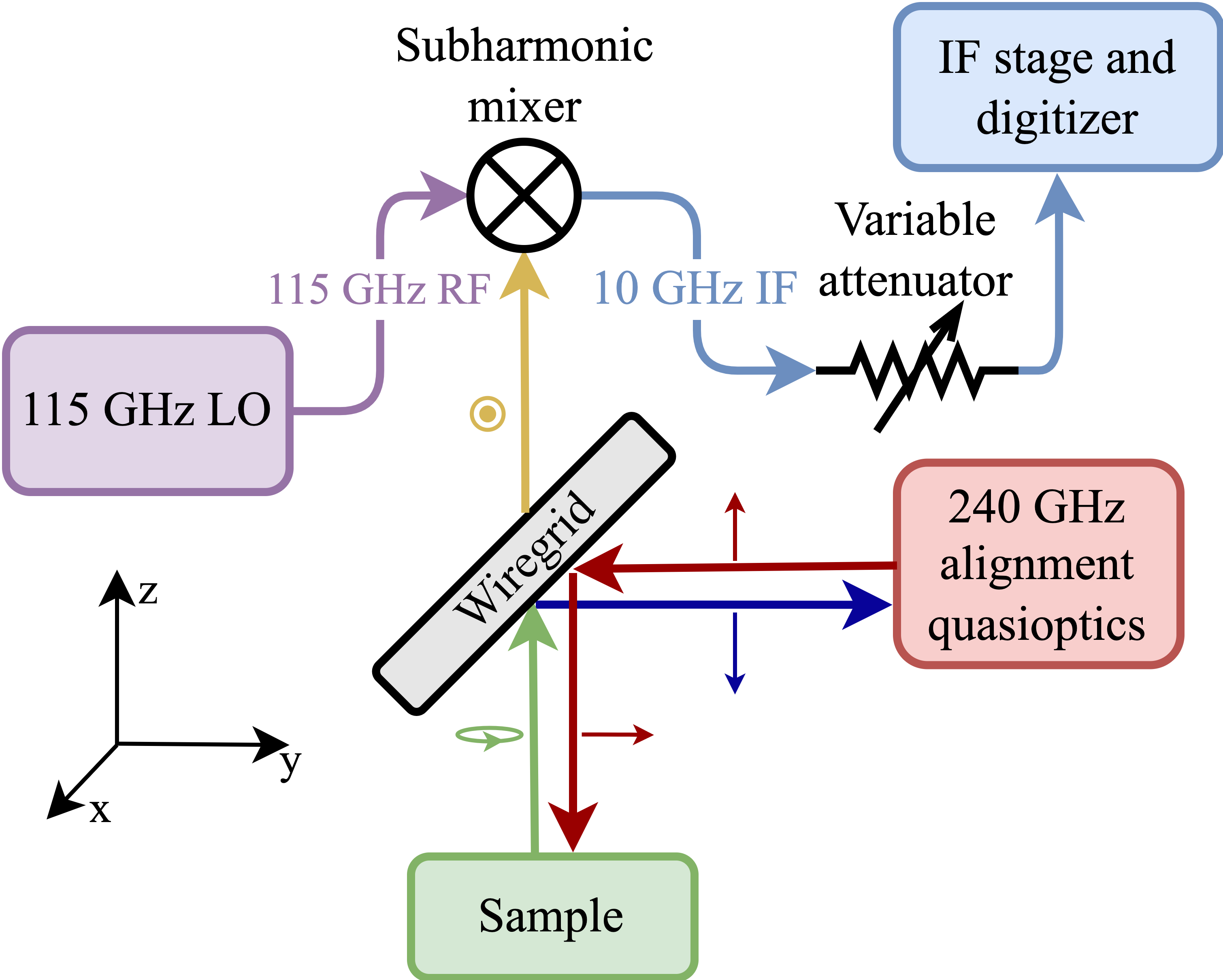}
\caption{Schematic drawing demonstrating ITST's heterodyne detection scheme for induction mode EPR.  $\vec{z}$-polarized 240\,GHz mw (red line) enter the waveguide through reflection in the vertical plane by a wiregrid polarizer (gray rectangle) and propagate along the waveguide $\vec{y}$-polarized (red line). Reflected mw are slightly elliptically polarized in the $xy$-plane (green line). The cross-polar $\vec{x}$-polarized mw (yellow line) are transmitted through the wiregrid polarizer and received by the WR3.4 Schottky-diode based subharmonic mixer (Virginia Diodes, Inc.) while the co-polar $\vec{y}$-polarization mw (blue line) are reflected by the wiregrid into an absorber. Cross-polar mw are mixed down to 10\,GHz (blue line) by the 115\,GHz local oscillator (purple rectangle and line). A variable attenuator limits the intermediate frequency (IF) power so that approximately -35\,dBm of 10\,GHz signal is transferred to the IF stage for filtering, amplification, additional mixing, and digitization. The polarization state of the beam in each path is depicted by small arrows.}
\label{fig:induction-mode}
\end{center}
\end{figure}

 ITST's spectrometer, which has been described in detail elsewhere \cite{takahashi_pulsed_2012, edwards_extending_2013}, operates in induction mode using a field-swept 12.5\,T Oxford Instruments\textcopyright{ }superconducting magnet. A custom, low power source is used (60\,mW, 240\,GHz; Virginia Diodes, Inc.) for cw or low-power-pulsed operation. Alternatively, ITST's free electron laser may be used as a source for kW power, $\mu$s-length pulses. A wiregrid polarizer at the end of the quasioptical bridge reflects  the polarized 240\,GHz radiation into a 1.2 m gold-plated corrugated waveguide that is coupled to the sample space. On resonance, the radiation that is reflected by the sample acquires a small, orthogonally-polarized component (``cross-polar") that is the EPR signal of interest. This signal is transmitted through the wiregrid at the end of the bridge, received by a WR3.4 Schottky-diode based subharmonic mixer (Virginia Diodes, Inc.), and subsequently mixed down to 10\,GHz  (see Fig. \ref{fig:induction-mode}). The 10\,GHz signal must be limited to no more than $-35$\,dBm in order to avoid saturating IF stage mixers and distorting the resulting EPR signal. Ideally, all the microwaves (mw) with polarization matching the incident radiation (``co-polar") are reflected back toward the source, and do not reach the receiver. However, without the rooftop mirror, enough co-polar radiation reaches the receiver that both the signal and co-polar leakage must be attenuated by a variable attenuator. This attenuation degrades the SNR, especially if the noise figure of subsequent IF amplifiers is greater than the noise of the attenuated IF signal. When utilizing the rooftop end mirror \cite{cruickshank_kilowatt_2009}, cross-polar isolation (the ratio of the desired cross-polar signal to the undesired co-polar signal) is greatly improved and the variable attenuator can be adjusted to attenuate much less IF power while still limiting it to $-35$\,dBm, resulting in improved SNR.

\subsection*{Design of Sample Holder}
\begin{figure}[htbp]
\begin{center}
\includegraphics[width=\linewidth]{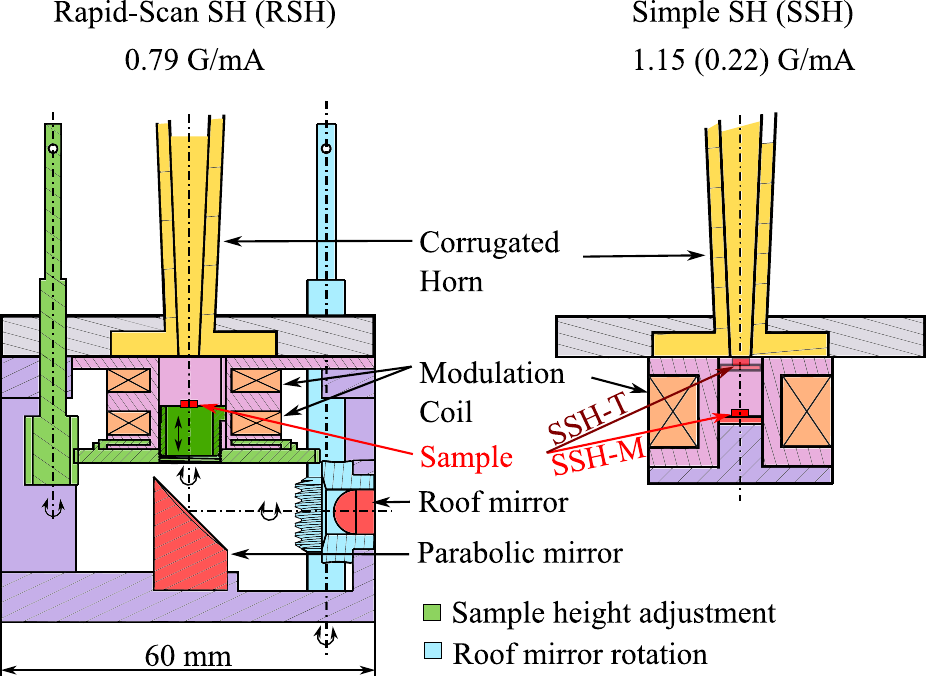}                     
\caption{Schematic drawing of quasioptical sample holder (QSH) and simple sample holder (SSH). The SSH has been used at ITST previously with two sample positions: the middle of the coil to maximize modulation amplitude (SSH-M), and the top of the coil to minimize losses due to divergence of the 240\,GHz Gaussian beam (SSH-T). The QSH employs two mechanical adjustment knobs, one to optimize sample position along the optical axis (green), and one to rotate the rooftop mirror to optimize cross-polar isolation (blue). The sample is positioned using a spur gear assembly; the large gear has 85 teeth and the small gear has 20. The rooftop mirror is rotated by a worm gear assembly with a 30:1 gear ratio.}
\label{fig:sample-holders}
\end{center}
\end{figure}

The quasioptical sample holder (QSH) was designed with a broad range of applications and user-friendliness in mind. In the case of the simple sample holder, which is similar to those used in many hfEPR spectrometers, the sample is placed close to the corrugated waveguide output at the top of the modulation coil (Fig. \ref{fig:sample-holders}, ``SSH-T") or on a mirror in the center of the modulation coil (Fig. \ref{fig:sample-holders}, ``SSH-M"). For pulsed EPR, the SSH-T position is used in order to minimize mw loss due to divergence of the incoming Gaussian beam. For rapid-scan and cwEPR, the SSH-M position is used to obtain the largest and maximally homogeneous modulation field at the sample space.

\begin{figure}[htbp]
    \centering
    \includegraphics[width=0.95\linewidth]{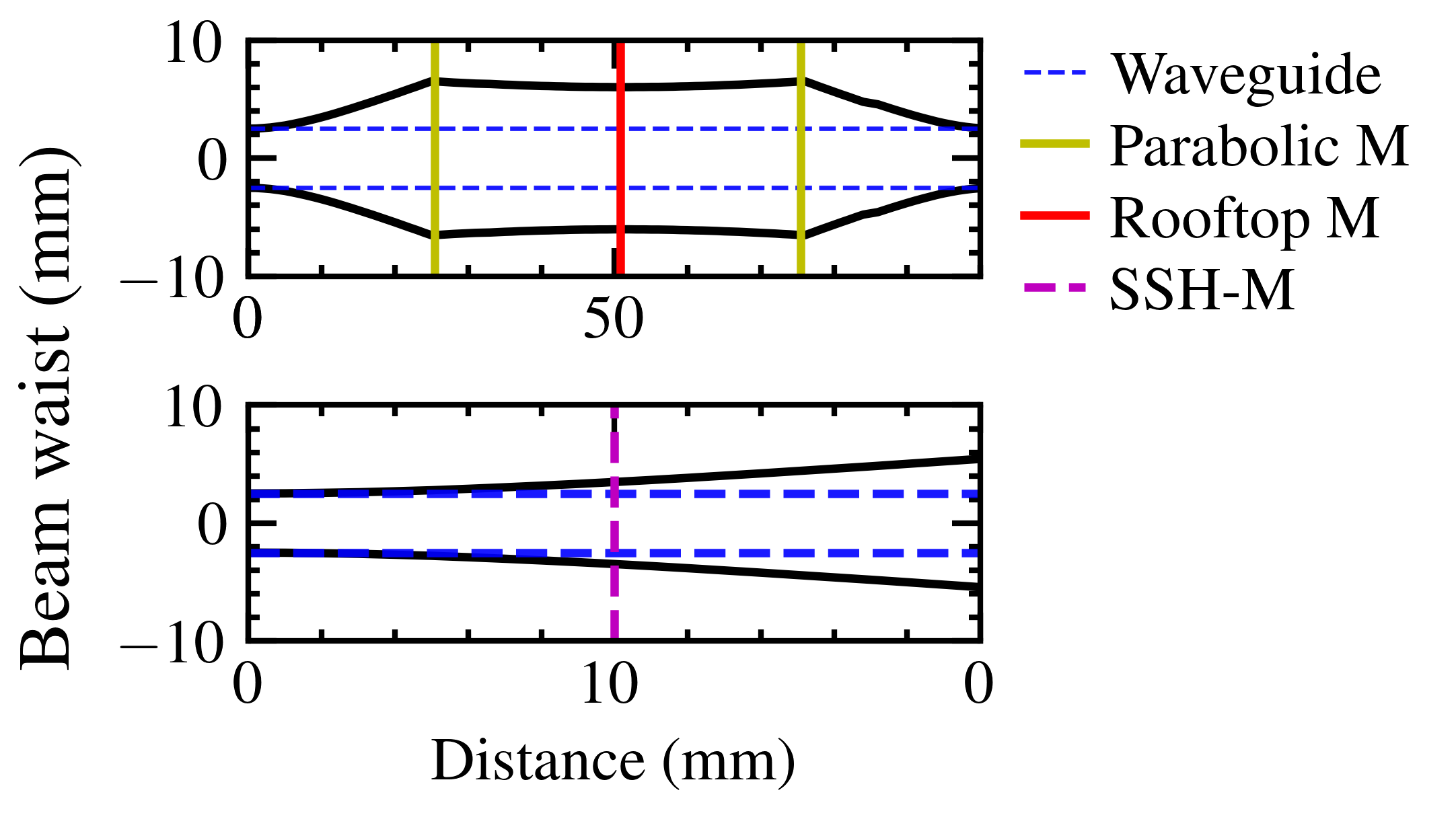}
    \caption{Simulation of Gaussian beam propagation through both sample holders \cite{goldsmith_quasioptical_1998}. The thin dotted blue lines represent the diameter of the input waveguide. The beams are reflected by a mirror at the middle of each propagation path. (Top) In the QSH, the beam expands after leaving the end of the waveguide (at $x=0$\,mm) and is focused by parabolic mirror (solid yellow lines) onto the roof mirror (at $x=51.8$\,mm, solid red line) and then back to the waveguide. (Bottom) In the SSH-M, after reflection from a flat mirror (at $x=10$\,mm, dashed purple line) the beam expands to a size larger than waveguide and results in microwave losses.}
    \label{fig:Beam_Waist}
\end{figure}

Our frequency-independent design for the QSH employs focusing quasioptics to ensure the beam waist is kept within the sample holder's specified optical path, reducing mw beam clipping due to the waveguide aperture. Additionally, the quasioptical design does not require a waveguide or end mirror within its modulation coil, which eliminates eddy currents inside the coil, and results in an improvement of field modulation amplitude and homogeneity. It is also capable of increasing cross-polar isolation and, therefore, reducing attenuation required for the IF stage which improves SNR. The QSH contains a $f=1"$ focusing parabolic mirror (MPD019-M03, Thorlabs Inc., USA) mounted along the axis of the corrugated wave guide. Inspired by previous work showing that precise, sub-degree rotation of a rooftop end mirror can result in 30\,dB of additional cross-polar isolation \cite{cruickshank_kilowatt_2009}, we have also mounted a machined Aluminium rooftop mirror at the focus of the parabolic mirror. A Gaussian beam-waist simulation \cite{goldsmith_quasioptical_1998} of the parabolic mirrors and rooftop mirror is shown in Fig. \ref{fig:Beam_Waist}.

The QSH also allows for positioning the center of the sample at a maximum of the magnetic field of the applied mw ($B_1$) through rotation of a spur gear assembly. The sample rests on 12.7\,$\mu$m-thick Mylar film that is glued to a slot-guided, threaded shaft. Rotation of the spur gear is translated into linear motion of the threaded shaft along the applied mw axis. 

The combination of increasing cross-polar isolation and maximizing $B_1$ results in an improvement of SNR, especially in the case of multilayer samples in which the sample bulk may lie at a magnetic field minimum and provide less-than-ideal SNR (\textit{e.g.} \cite{wilson_adventures_2019, maity_triggered_2023}). Translation along the mw axis is not necessary in the case of the SSH, as the standing wave condition ensures that an electric field node (and magnetic field antinode) appears on the surface of the mirror. 

The QSH was designed for 3D-printing; the testing prototype and gears were 3D-printed from polylactic acid (PLA) by an Original Prusa i3 MK3S+ (Prusa Research s.r.o., CZ). The sample holder was attached to the EPR probe by screws; all screw threads were tapped post-print. The holder was initially designed for room temperature experiments, but with a different printing substrate (\textit{e.g.}, nylon-based filament), liquid nitrogen temperatures should be easily achievable. 

The modulation coil was wound from Cu 32 AWG wire. We employed a Helmholtz modulation coil to get direct access to the sample space. Sample space access is necessary for optical excitation in experiments similar to those presented in Ref. \cite{maity_triggered_2023} and may also allow for fast sample loading in future designs. The coil was calibrated using a reference sample of LiPc (see S.I. Fig. \ref{fig:Modulation}).

\section{Experimental Results}
The QSH was tested with three different samples and EPR modes (see Table \ref{T:S/N-Ration}): a solution of 100\,$\mu$M Gadolinium chloride in D$_2$O by cwEPR, a diamond crystal with nitrogen substitutional defects (P1 centers) by pulsed EPR, and a single crystal of LiPc by rapid-scan EPR. The measurements were compared with those done with the SSH, in both the SSH-M and SSH-T positions. The QSH experiments were done with a flat (QSH-FM) and rooftop end mirror (QSH-RM) for comparison.

\subsection{Continuous wave EPR}
\begin{figure}[htbp]
\begin{center}
\includegraphics[width=0.8\linewidth]{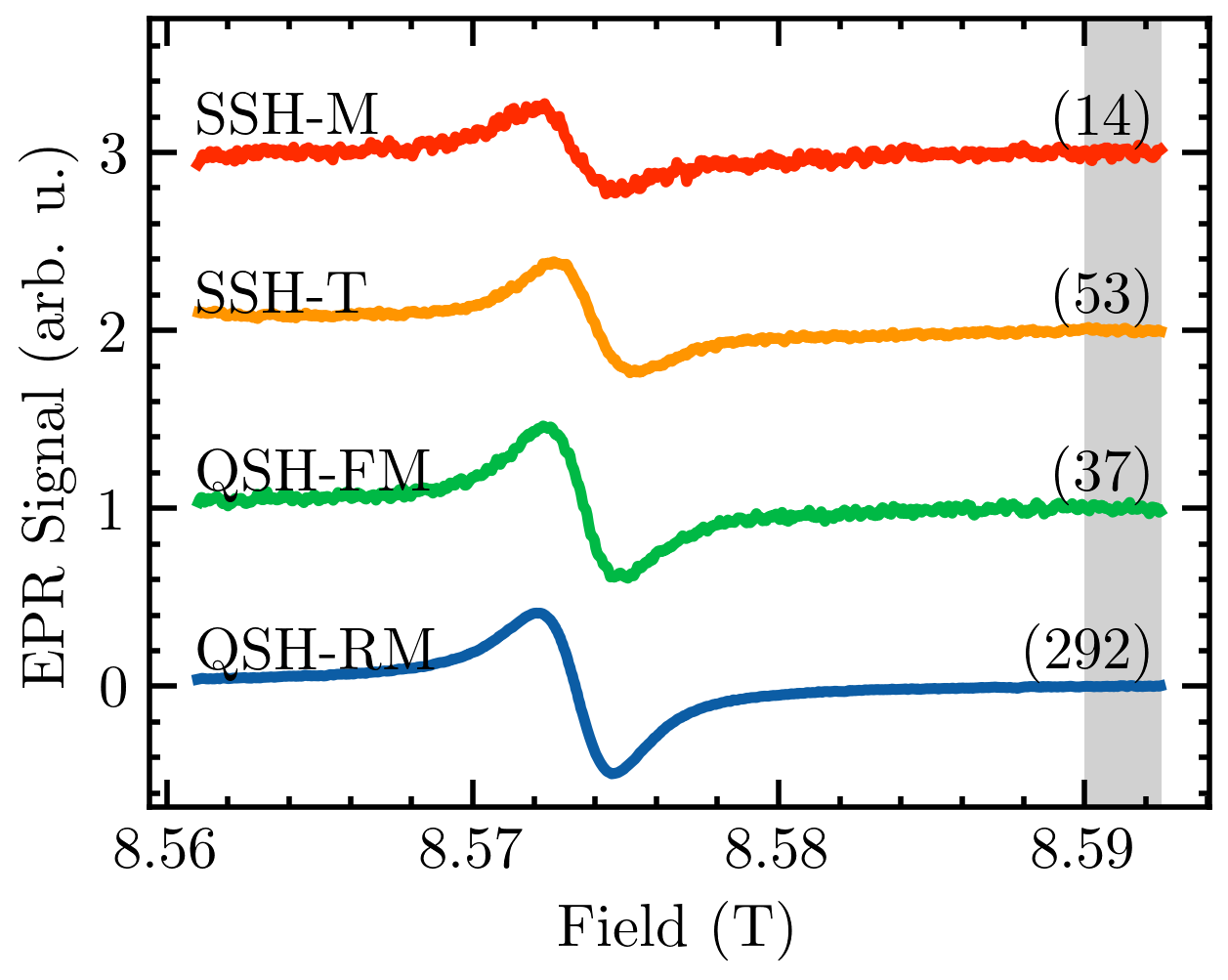}                     
\caption{Continuous wave EPR of 100 $\mu$M GdCl$_3$. Lock-in amplifier (100 ms lock-in time constant) detected signal intensity as a function of field. Applied magnetic field was modulated at the resonant frequency of the modulation coil and accompanying 10\,nF capacitor ($B_{mod}=14.8$\,G at $33$\,kHz in the case of the SSH, $B_{mod}=7.9$\,G at $23$\,kHz in the case of the QSH). The SNR was best in the case of the QSH-RM because of the ability to reduce co-polar noise, as well as to position the thin GdCl$_3$-filled capillary at a maximum of the irradiated $B_1$ field. The baseline was removed by subtracting the mean of points within the vertical gray column. SNR (in parentheses) was calculated by dividing peak baseline-corrected signal by the standard deviation of baseline-corrected points within the vertical gray column.}
\label{fig:CWEPR}
\end{center}
\end{figure}

cwEPR was performed on 100\,$\mu$M GdCl$_3$ dissolved in D$_2$O. The GdCl$_3$ solution was inserted into a 100\,$\mu$m-thick, 2-by-5\,mm borosilicate glass capillary (VitroCom, Mountain Lakes, NJ) in order to maximize the ratio of surface area to thickness, and sealed with wax \cite{wilson_adventures_2019}. The modulated signal was amplified using a lock-in detector (SR830, Stanford Research Systems) and is shown in Fig. \ref{fig:CWEPR}. The SNR was best in the case of QSH-RM ($\text{SNR}=292$) because of our ability to minimize the co-polar baseline while also maximizing the spin response by optimizing the sample position. Next best was SSH-T ($\text{SNR}=53$), due to the near-zero mw losses as a result of the mirror being placed directly at the end of the waveguide. SSH-M had the lowest SNR ($\text{SNR}=14$): this is due to mw losses as a result of divergence of the Gaussian beam. QSH-FM ($\text{SNR}=37$) had an SNR greater than that of SSH-M because it was able to reduce divergence loss with focusing quasioptics, but, as expected, still had more divergence loss than SSH-T, which had no free-space propagation.

\subsection{Pulsed EPR}
\begin{figure}[htbp]
\begin{center}
    \includegraphics[width=0.8\linewidth]{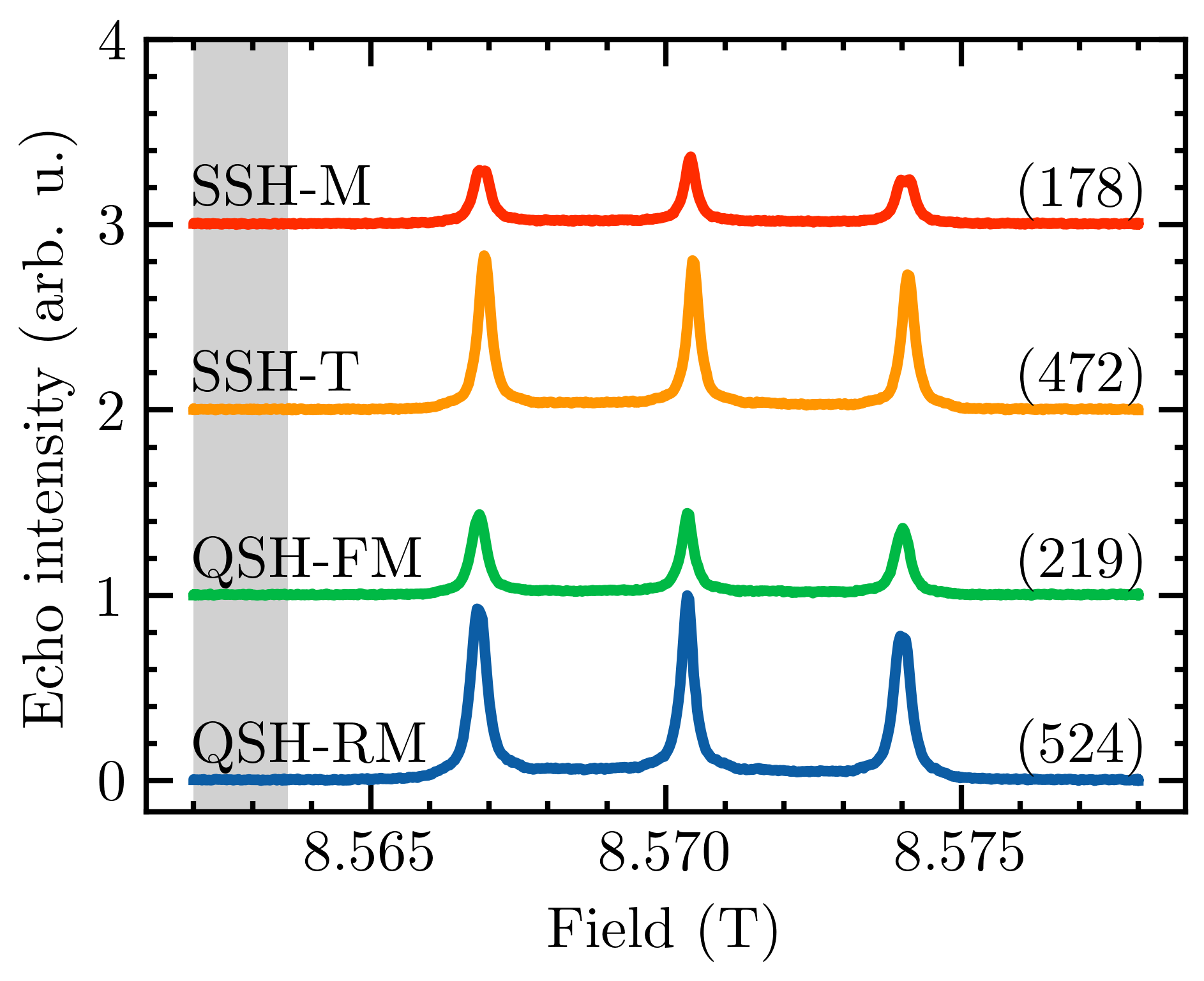}                 
    \caption{Field-swept echo of P1 centers in diamond. Integrate echo magnitudes plotted as a function of field for each sample holder configuration (1\,kHz repetition rate with 512 averages at each field position). SSH-M (red) performed the worst ($\text{SNR}=178$). SSH-T (orange) performed quite well---the optical path length was minimal and therefore so were losses due to propagation ($\text{SNR}=472$). QSH-FM (green) performed second worst ($\text{SNR}=219$). QSH-RM (blue) performed the best as a result of focusing optics and the ability to maximize desirable cross-polar signal ($\text{SNR}=524$). SNR was calculated by dividing peak signal by the standard deviation of points within the vertical gray column.}
    \label{fig:pulsed-results}
\end{center}
\end{figure}

Pulsed EPR was performed at 240\,GHz on P1 nitrogen vacancy centers in a diamond crystal using a cw 60\,mW Virginia Diodes Inc. source that was pulsed using a fast PIN diode switch (SWM-0JV-1DT-2ATT, American Microwave Corporation, USA). A Hahn echo pulse sequence, consisting of a 600\,ns excitation pulse, 1.2\,$\mu$s delay time, and 800\,ns refocusing pulse was used to create a spin echo. The pulse sequence was repeated at 1\,kHz and averaged over 512 repetitions per field position. Integrated echo intensity was recorded as a function of field and plotted in Fig. \ref{fig:pulsed-results}. The results from each sample holder were quite similar. The SSH-M performed the worst ($\text{SNR}=178$) because its long optical path length and beam divergence cause clipping at the waveguide aperture (see Fig. \ref{fig:Beam_Waist}). The SSH-T did not suffer from this clipping, and therefore performed second best ($\text{SNR}=472$). The QSH-FM ($\text{SNR}=219$) and the QSH-RM ($\text{SNR}=524$) should have performed similarly, as the co-polar background cannot obscure the cross-polar signal after the pulse sequence has occurred. However, optimizing the roof mirror angle for the maximum Hahn echo improved SNR and allowed the QSH-RM to perform the best overall.

\subsection{Rapid-scan EPR}

Rapid-scan EPR was performed on a needle of LiPc at 240\,GHz in three sample sample holders (SSH-M, QSH-FM, QSH-RM). Rapid-scan SSH-T experiments were not performed because the modulation field strength (approx. 16\,G tip-to-trough, see S.I. Fig. \ref{fig:Modulation}) was insufficient to entirely sweep the LiPc resonance. In order to maximize the improvements, in the case of the QSH-FM, the sample position was varied until the signal-to-baseline ratio was maximized. For the QSH-RM, a two-step optimization was performed: first, with the magnetic field tuned to resonance, the sample position was varied until the signal-to-baseline ratio was maximized; next, with the magnetic field off resonance, the rooftop mirror was rotated to minimize the power received by the induction mode detector. This procedure minimized the baseline and maximized the ratio of cross-polar to co-polar signal. The rooftop mirror increased the cross-polar isolation from $\sim$30\,dB to $\gtrsim50$\,dB.

\begin{figure}[htbp]
\begin{center}
\includegraphics[width=\linewidth]{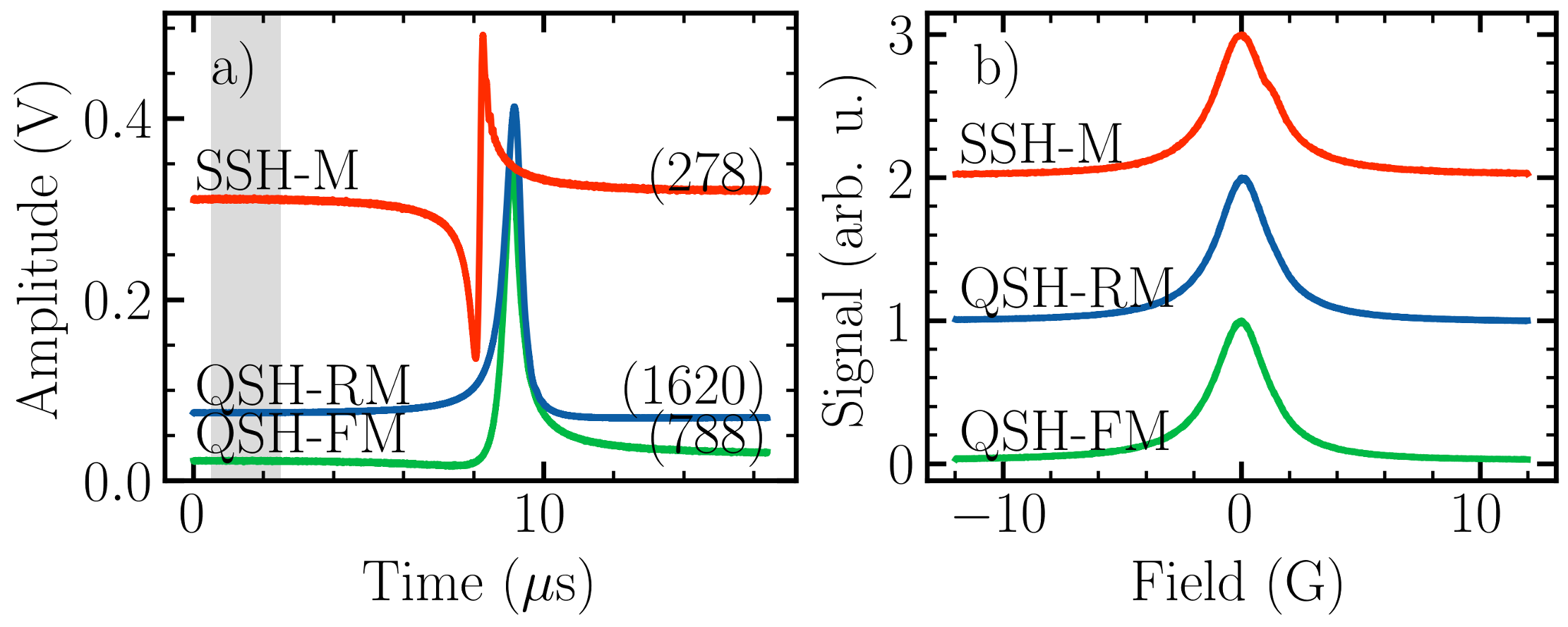}                   
\caption{Rapid-scan EPR of LiPc needle, directly detected and averaged. (a) Time-resolved rapid-scan EPR spectra of LiPc with 10,000 averages. SSH-M (blue) ($B_{mod}=144$\,G at $33$\,kHz, $A_{\text{var}}=25.5$\,dB, $\text{SNR}=278$); QSH-FM (green), sample position to achieve maximum resonant signal amplitude was adjusted using optical axis positioning thread ($B_{mod}=75$\,G at $23$\,kHz, $A_{\text{var}}=20.5$\,dB, $\text{SNR}=788$); QSH-RM (red), sample position was adjusted to achieve maximum resonant signal amplitude and baseline was minimized with rooftop mirror rotation ($B_{mod}=75$\,G at $23$\,kHz, $A_{\text{var}}=5.75$\,dB, $\text{SNR}=1620$). SNR was calculated by dividing peak amplitude by the root-mean-squared deviation of points within the vertical gray bar. (b) Deconvolved LiPc spectra after application of the driving-function Fourier deconvolution algorithm found in Ref. \cite{tseytlin_general_2020}.}
\label{fig:rs-results}
\end{center}
\end{figure}

As shown in Table \ref{T:S/N-Ration} and Fig. \ref{fig:rs-results}, the SNR of the QSH represented an improvement over the SSH. In the QSH-FM, the SNR improved by approximately $3\times$ over the SSH ($278\rightarrow788$). After replacing the end mirror with a rooftop mirror (QSH-RM) the SNR was improved by an additional $2\times$ ($788\rightarrow1620$), for a total improvement of $6\times$ ($278\rightarrow1620$). 

\section{Discussion}
The QSH showed SNR improvements for all three EPR experimental techniques: cw, rapid-scan, and pulsed. In the case of cw and rapid-scan EPR, reducing the baseline is critical for improving the SNR. In this case, the rooftop mirror provides a significant benefit. It is capable of improving co-polar isolation by $\gtrsim20$\,dB (from $\sim$30\,dB to $\gtrsim50$\,dB) and SNR by around 6, when compared to the best possible configuration using the SSH. It is worth noting that the results presented here were established using a rooftop mirror end-milled from an aluminum rod. Precise finishing or gold-coating the rooftop surface may provide improved reflectivity and therefore reduced scattering; this may allow for even greater co-polar isolation and SNR improvements. Additionally, the QSH benefited cw and rapid-scan by allowing the sample to be placed in the middle of the modulation coil where the best possible performance could be achieved. 

In pulsed EPR, the signal is usually detected after the excitation is turned off, which makes co-polar leakage a smaller concern than in the othe two modes. Therefore, in this mode, the sample is placed as close as possible to the waveguide to limit losses due to Gaussian beam propagation. Such losses cannot be avoided in the QSH-FM/RM, but they can be partially mitigated by the parabolic focusing mirrors. Nevertheless, even for the Hahn echo experiment presented here, the QSH-RM was able to improve SNR by $\sim$15\%. In order to achieve this improvement, the polarization of the echo was slightly rotated to better couple with the induction mode quasioptics, and the sample was positioned at a magnetic field maximum.

Future work for the QSH-RM includes improving the reflectivity of the end-milled rooftop mirror, incorporating a fast-loading sample design, and identifying and optimizing a new 3D-printing substrate that is capable of handling liquid nitrogen temperatures for cryogenic experiments. It may also be beneficial to implement a piezo-electric rotator in place of the manual rotation knob in order to optimize sample position and rooftop mirror angle repeatably and automatically. Further, the QSH may provide an avenue to greatly reduce dead-time of pulsed EPR using high power mw sources such as ITST's free electron laser (FEL).

\begin{table}[htb]
 
    \centering 
    \caption{\label{T:S/N-Ration}SNR and insertion loss for each SH config.}
    \begin{tabular}{l|rrrr}
     SH config.     & CW        & Pulsed     & RS    & Ins. loss (dB)\\
     \hline
     Sample         & Gd(III)   & P1 Dia. & LiPc  & --\\
     SSH-M          & 14        & 178        & 278   & 4.18$\pm$0.2\\
     SSH-T          & 53        & 472        & --    & 2.67$\pm$0.4 \\
     QSH-FM        & 37   &  219  & 788  & 1.84$\pm$0.6\\
     QSH-RM        & 292  &  524 & 1620  & 1.53$\pm$0.4\\
    \end{tabular}

\end{table}

\section{Sample details}

\subsection*{P1 diamond}

The diamond sample was a 5mm-by-5mm-by-1mm type 1b diamond crystal with substitutional nitrogen (``P1") defects. Estimated P1 concentration is approximately 62$\pm$5 ppm \cite{wilson_adventures_2019}. The crystal was adhered to its mount using Apiezon\textcopyright{ }vacuum grease (sample mount was a flat mirror in the case of the simple SH, and a 12.7 $\mu$m piece of Mylar\textregistered{ }film in the case of the QSH).

\subsection*{Gadolinium(III) chloride}

Gadolinium chloride was bought from Sigma Aldrich and dissolved in D$_2$O with a concentration of 1\,mM and then diluted to 100\,$\mu$M. The spectra shown in Fig. \ref{fig:CWEPR} were recorded without sample degassing. The capillary rested on its mount and stayed in position due to gravity (sample mount was a flat mirror in the case of SSH, and a 12.7 $\mu$m piece of Mylar\textregistered{ }film in the case of the QSH).

\subsection*{Lithium phthalocyanine (LiPc)}

LiPc in micro crystalline form was prepared electrochemically following a procedure described in the literature \cite{turek_preparation_1987,afeworki_preparation_1998}. A crystal sample was obtained from Mark Tseytlin (West Virginia University). A sharp single needle was adhered to its mount using Apiezon\textcopyright{ }vacuum grease (sample mount was a flat mirror in the case of the simple SH, and a 12.7 $\mu$m piece of Mylar\textregistered{ }film in the case of the QSH).

\section{Conclusion}
In this paper, we demonstrated a rapidly prototyped, cost-efficient 3D-printed sample holder that is capable of greatly improving the SNR of cw and rapid-scan EPR by reducing co-polar leakage and optimizing sample position to a $B_1$ antinode. Co-polar isolation was improved by a factor of $\gtrsim20$\,dB and SNR was improved by approximately 6, representing a 36$\times$ reduction in signal acquisition time for comparable SNR. For cw or rapid-scan experiments with low spin concentrations, a reduction in required acquisition time may enable experiments that are otherwise inaccessible.

The sample holder presented here represents an easy-to-implement solution for improving the SNR of variety of homebuilt EPR spectrometer designs through custom-fit modifications of the .stp files included in the S.I. 3D-printed components enable simple incorporation into a variety of induction-mode hfEPR spectrometers and may provide a method to improve SNR in the field.

\section*{Acknowledgments}
The authors would like to thank Dr. Nikolay Agladze for his thoughtful insight regarding quasioptical design and his assistance with Gaussian beam propagation simulations.

\ifCLASSOPTIONcaptionsoff
  \newpage
\fi

\bibliographystyle{IEEEtran}
\bibliography{references_manual}

\begin{IEEEbiography}[{\includegraphics[height=1.25in,clip,keepaspectratio]{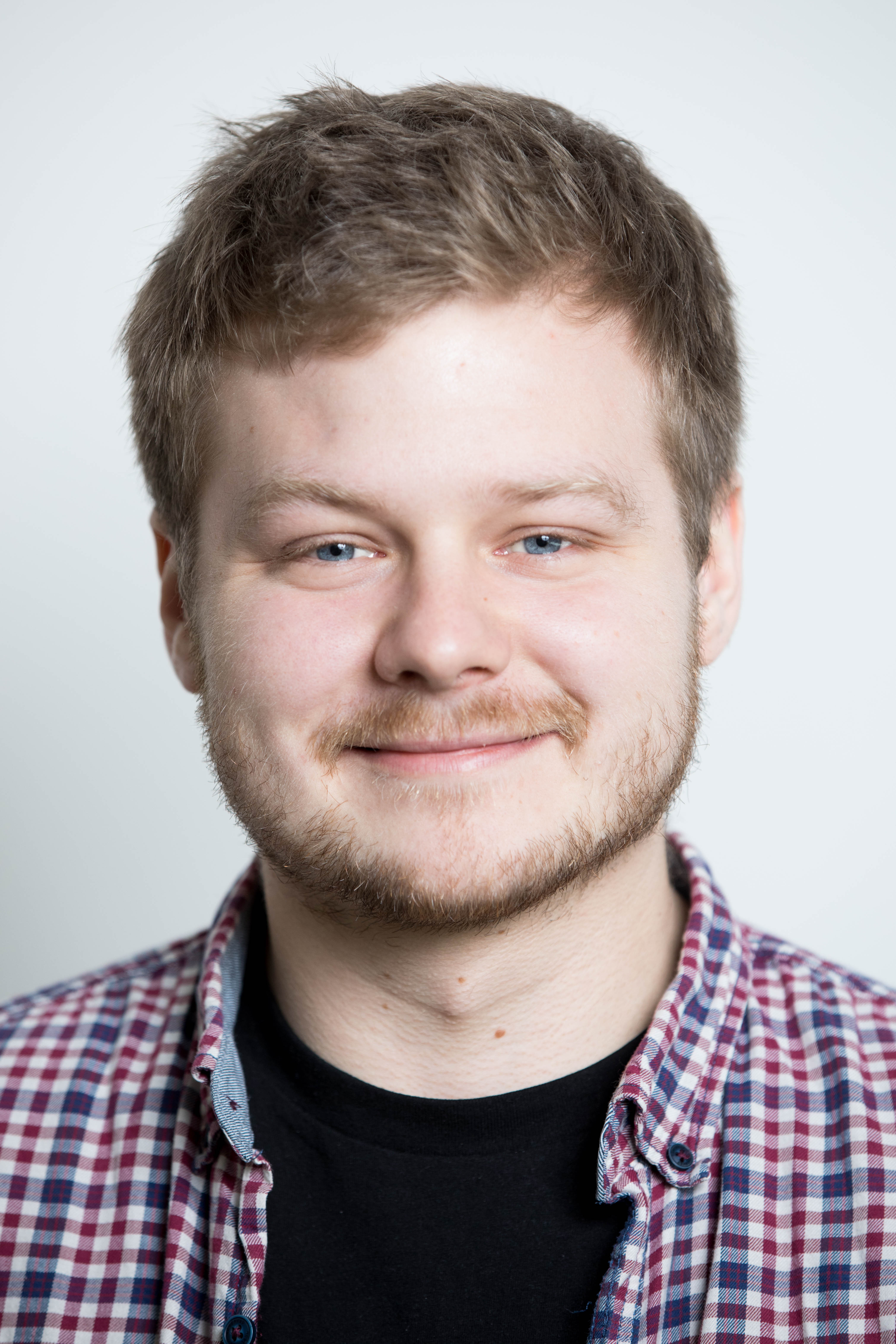}}]%
{Antonin Sojka}
received his Bachelor's and Master's degrees from the Brno University of Technology, Czech Republic in 2016 and 2018. During this time, he designs and develop a novel Ultra High Vacuum Scanning Tunneling Microscope for in-situ measurement of molecule deposition of a variety of surfaces. He did his Ph.D. at the Central European Institute of Technology BUT since 2018 under the supervision of Petr Neugebauer in the group Magneto-Optical and Terahertz Spectroscopy (MOTES), where his main research interest is the development of High Field broad frequency ESR for relaxation studies with focused on rapid-scan technique. At the moment he is working on the development of new high-field EPR spectrometer power by FEL at the University of California, Santa Barbara.   
\end{IEEEbiography}
\begin{IEEEbiography}[{\includegraphics[width=1in,clip,keepaspectratio]{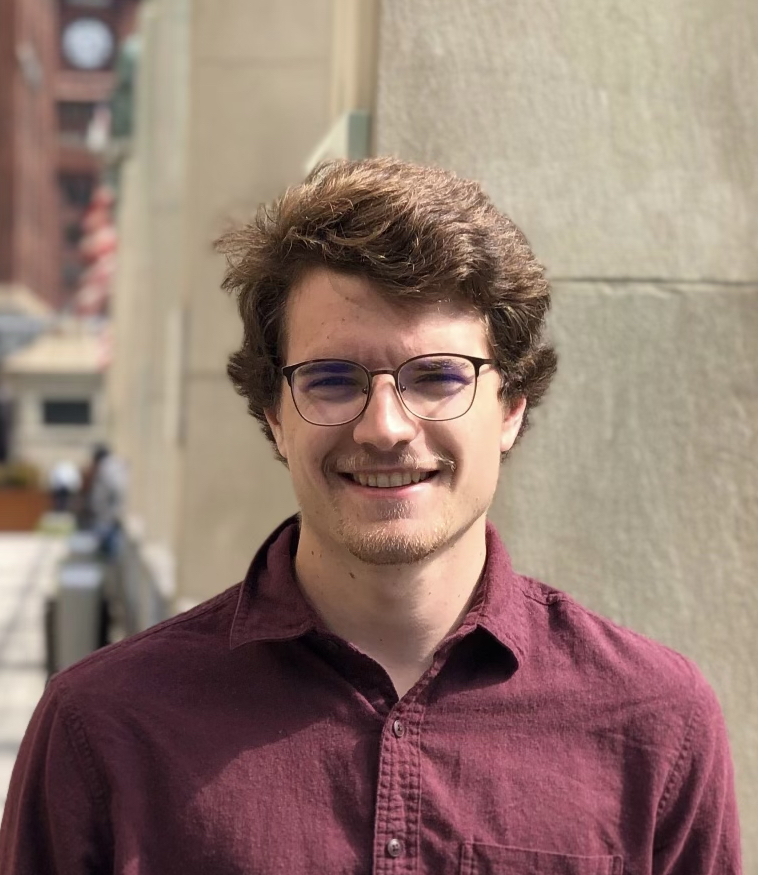}}]%
{Brad D. Price}
received his BSc in Physics from the University of Northern British Columbia in 2019. He worked on developing techniques for non-destructive analysis of engineered wood products, as well as 3D-printed optics for sub-THz Gaussian-to-flat-top beam shaping under Dr. Matt Reid. He received his MA in Physics from the University of California, Santa Barbara in 2022 for his work on filming triggered functional dynamics of light-activated proteins using high-field EPR under Dr. Mark Sherwin. He is currently continuing his work on protein filming as well as developing high-field EPR instrumentation and techniques, which include free-electron laser and rapid-scan EPR. 
\end{IEEEbiography}

\begin{IEEEbiography}[{\includegraphics[width=1in,clip,keepaspectratio]{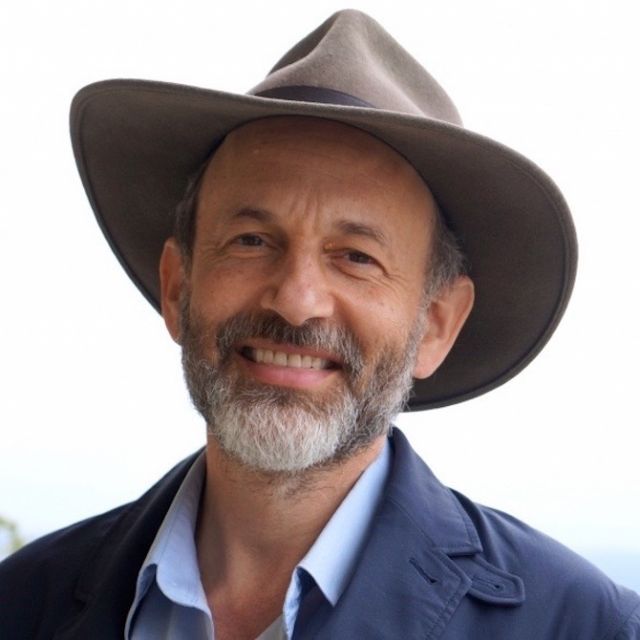}}]
{Mark S. Sherwin} received his Ph.D. degree in experimental solid state physics from the University of California (UC), Berkeley, CA, USA, in 1988. He then joined the faculty of the Physics Department at UC Santa Barbara, CA, USA. He is the Director of the Institute for Terahertz Science and Technology at UC Santa Barbara, the home of the UC Santa Barbara Free-Electron Lasers. His research group focuses on driven quantum and biological matter. Current interests include terahertz mixers, the nonlinear response of semiconductors to intense terahertz radiation, high-field pulsed electron paramagnetic resonance, and the triggered functional dynamics of proteins. Dr. Sherwin is a Fellow of the American Physical Society.
\end{IEEEbiography}

\end{document}


\title{\textbf{Supplementary information:} Order-of-Magnitude SNR Improvement for High-Field EPR Spectrometers via 3D-Printed Quasioptical Sample Holders}

\author{Antonín~Sojka,
        Brad~D.~Price,
        Mark~S.~Sherwin
}

\maketitle
\IEEEpeerreviewmaketitle

\section{QSH sample holder details}

\subsection{Rooftop mirrors}
One benefit of the quasioptical design is spatial separation between the sample space and end mirror. As a result of this additional space, a rooftop mirror was easily incorporated into the SH. Rooftop mirrors are well-known polarization rotators \cite{latmiral_radar_1962, challener_achromatic_1996, smith_quasi-optical_1998, budil_jones_2000, blok_continuous-wave_2004, chuss_interferometric_2006, schillaci_effect_2012}, and are able to significantly reduce unwanted co-polar noise in induction mode EPR spectrometers \cite{smith_high-field_2008}: many high-field EPR spectrometers operate in induction mode (see Fig. \ref{fig:induction-mode}), meaning only polarization orthogonal to the irradiated polarization is of-interest and detected by the mw receiver \cite{teaney_microwave_1961, fuchs_high-fieldhigh-frequency_1999}. Rooftop end mirrors are capable of reducing co-polar signal in induction mode by correcting for small polarization rotations from imperfect couplings throughout the quasioptical setup as well as transmission through the sample itself. 

The Jones matrix for a roof mirror is
\begin{equation}
    \textbf{R}(\theta)=
    \left( 
    \begin{array}{cc}
        \cos2\theta & \sin2\theta \\
        -\sin2\theta & \cos2\theta \\
    \end{array} 
    \right)
\end{equation}
where $\theta$ is the angle between the incident polarization and the axis of the roof mirror \cite{chuss_quasioptical_2005}. This matrix demonstrates the $2\theta$ polarization rotation due to an angular offset from the y axis, $\alpha$, as shown in the xy basis:
\begin{align*}
    \textbf{R}(\theta)\vec{E}&=
    \left( 
    \begin{array}{cc}
        \cos2\theta & \sin2\theta \\
        -\sin2\theta & \cos2\theta \\
    \end{array} 
    \right)
    \left( \begin{array}{cc} x \\ y \end{array} \right) \\
    &=
    \textbf{R}(-\alpha/2)
    \left( \begin{array}{cc} \sin\alpha \\ \cos\alpha \end{array} \right) \\
    &=       \left( 
    \begin{array}{cc}
        0 \\
        1 \\
    \end{array} 
    \right)
\end{align*}
which has the effect of rotating all co-polar radiation into the direction that is discarded by the y-axis induction mode wiregrid. 

\subsection{3D-printing-optimized design}
\begin{figure}[htbp]
\begin{center}
\includegraphics[width=\linewidth]{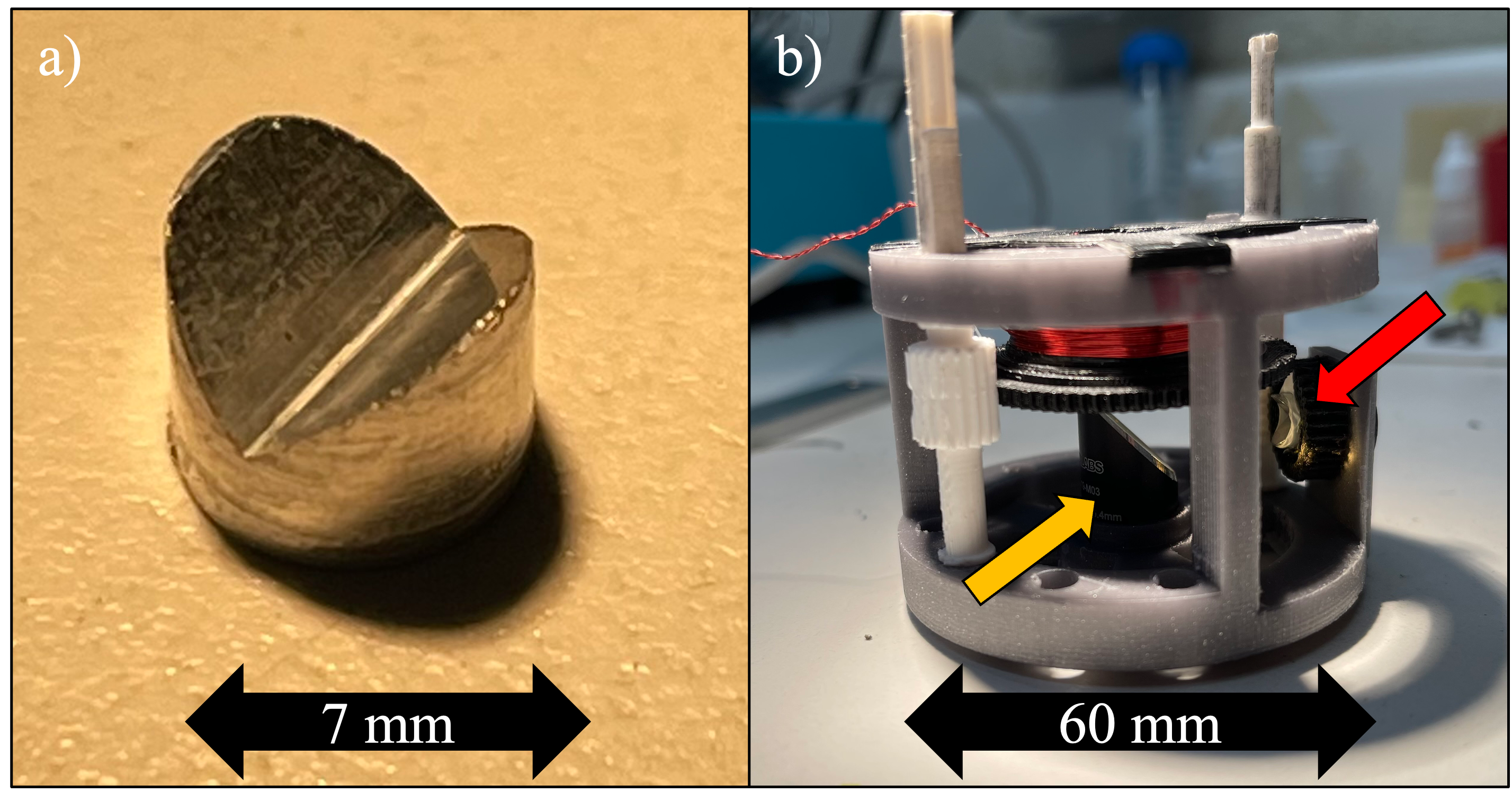}   
\caption{Rapid-scan sample holder components. (a) Polarization-rotating rooftop mirror end-milled from 7 mm aluminum rod. (b) Rapid-scan sample holder assembly. All parts except the 1/2" parabolic mirror (orange arrow) and rooftop mirror (red arrow) were 3D-printed for fast prototyping and low cost implementation.}
\label{fig:roof-mirror}
\end{center}
\end{figure}

This section discusses the assembly of the sample holder with on how to print the 3D model. Print settings, as discussed here, were optimized for use on a Prusa MK3+ 3D-printer. The final assembly is shown in Fig. \ref{fig:roof-mirror}, and can be found as 3D step file as additional supplementary file.

All parts (except the parabolic mirror and the roof mirror) were printed from the 1.75 mm PLA filament Prusament (Prusa Research s.r.o., CZ). To achieve a rigid printed structure, the model was printed vertically with supports with z-step 0.1 mm. The default printer settings for PLA print were used.

Aside from the assembly shown in Fig. \ref{fig:roof-mirror}, two additional rigid rotation rods were required to allow the user to transmit a rotation from outside the magnet bore to the sample space. The sample holder shown was designed for 60 mm magnet bore, but small modifications would allow it be used in a smaller bore. The only parts that needed to be purchased were the gold-coated parabolic mirror, the coated copper wire, and the Mylar film. The Mylar film could be replaced by other mm-wave transparent materials, depending on use case. Thin Mylar was used here to minimize scattering, absorptive loss, and standing waves.

\subsection{QSH mechanical tunability}
\begin{figure}[htbp]
\begin{center}
\includegraphics[width=\linewidth]{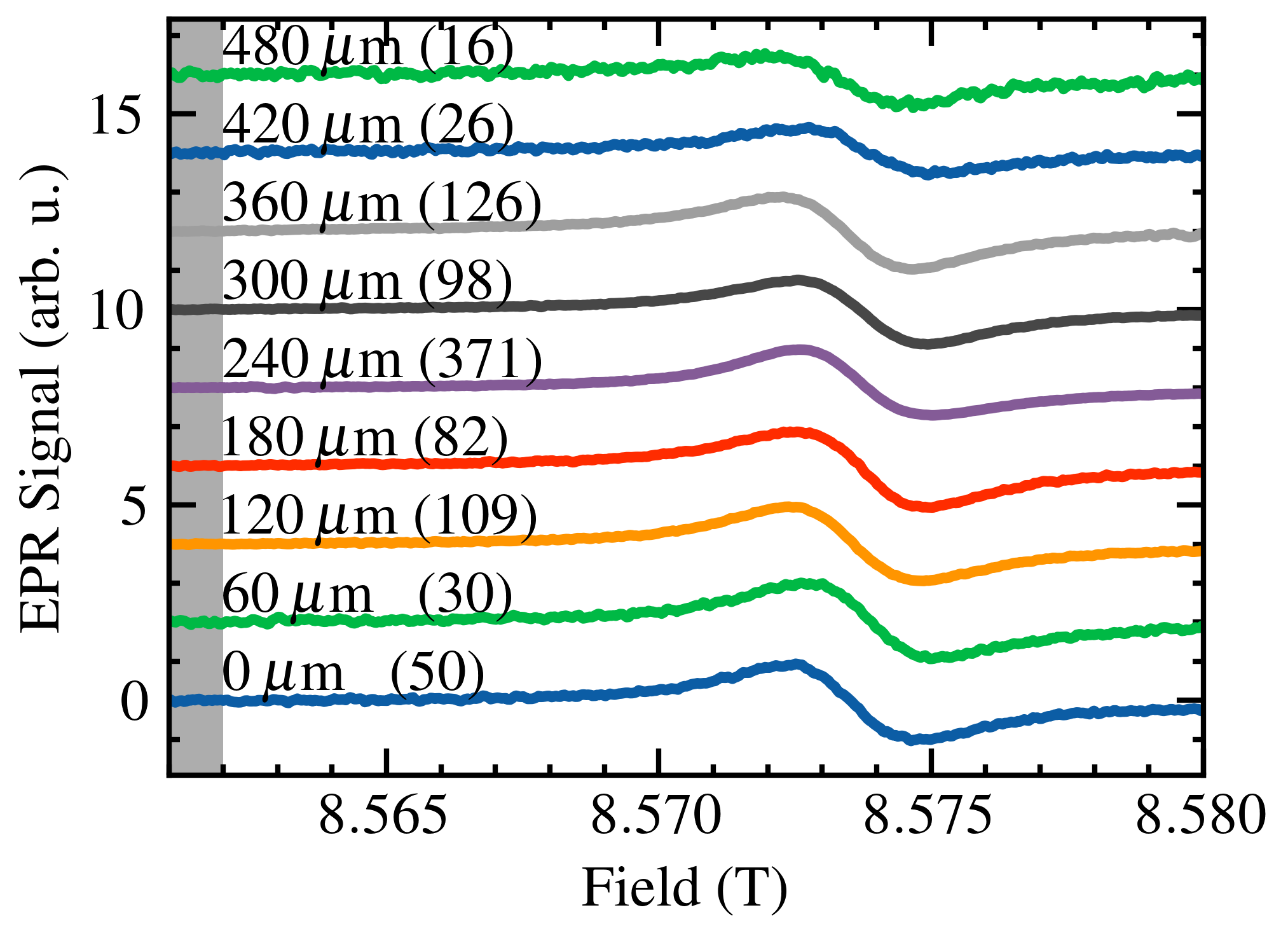}   
\caption{Measurement of signal dependence on sample position in the sample holder with a flat end mirror. 100$\,\mu$M Gd dissolved in D$_2$O was placed in a thin, 2-by-5 mm borosilicate glass capillary (VitroCom, Mountain Lakes, NJ) on Mylar film (12.7 $\mu$m thick) in the middle of modulation coil $z=0\,\mu$m. In steps of 6$\,\mu$m, measurements were performed. Lineshapes at each position are shown along with their respective SNRs in parentheses. SNR was calculated by dividing peak signal by the standard deviation of points within the vertical gray column.}
\label{fig:Sample_Position}
\end{center}
\end{figure}

There are two mechanical degrees of freedom implemented into the RSH holder: motion along the optical axis and rotation of the rooftop mirror. Sample positioning along the optical axis ($z$-direction) is crucial for thick or multilayer samples as the user is able to ensure that B$_{ext}$ is maximized at middle of the sample of interest. Fig. \ref{fig:Sample_Position} shows that for the thin liquid sample in capillary, sample positioning may be increase SNR by as much as 24$\times$. However, the two degrees of freedom are not completely independent, and changing one alters the optimal position for the other -- changing the sample position can slightly modify the output polarization and changing the angle or position of the imperfect rooftop mirror can alter the locations of $B_{ext}$ antinodes. This makes signal optimization tricky, but possible, through trial-and-error. 

\begin{figure}[htbp]
\begin{center}
\includegraphics[width=\linewidth]{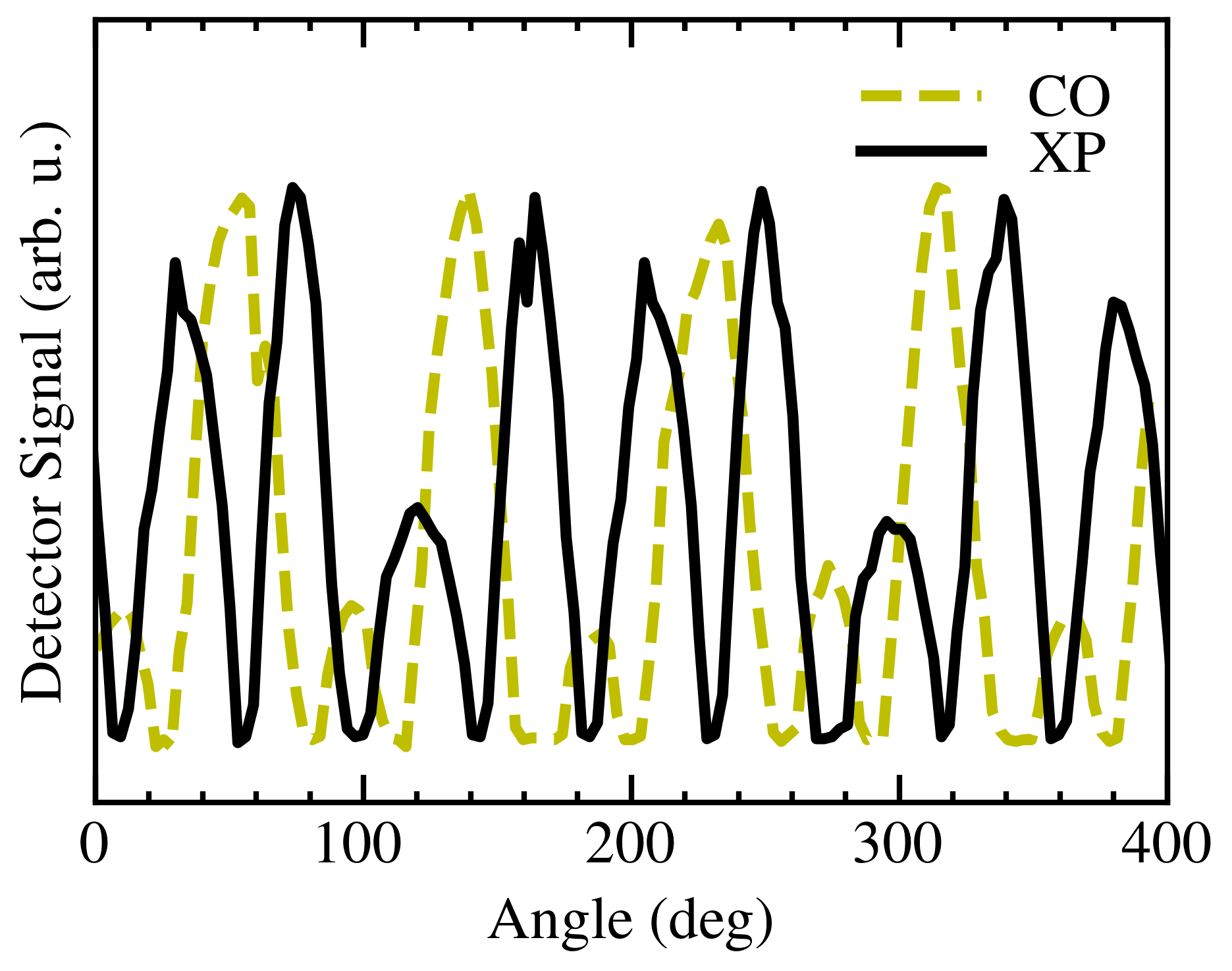}   
\caption{Normalized measurement of the power dependence upon rotation of roof mirror in co-polar (CO) and cross-polar (XP) signal. The measurements were with the ACST THz detector placed in the co- and cross-polar EPR detector positions. In the ideal case, the position that transmits minimal power to the co-polar line should transmit maximum power to the cross-polar line and vice versa. Because the rooftop mirror had to be rotated by hand in this prototype, angle step errors were introduced and some maxmima and minima do not overlap exactly.}
\label{fig:copolar}
\end{center}
\end{figure}

The second degree of freedom employed a worm gear system to rotate a rooftop end mirror that was perpendicular to the waveguide axis. Upon reflection, polarization may be rotated without loss. The user aims to reduce co-polar power leaking from source to detector and thereby increase the cross-polar isolation in induction mode. In Fig. \ref{fig:copolar}, we measured dependence of the co-polar signal intensity upon rotation of the rooftop mirror with a quasioptical THz Detector (3DL 12C LS2500 A2, ACST gmbH, DE). From the measurements, it is visible that at certain position, mw power can be in cross- or co-polar almost entirely. Compared to the flat mirror, we were able to achieve greater than 20 dB of extra cross-polar isolation.

\section{Modulation}\label{sec:calibration}
The calibration value (G/mA) for the modulation coil is crucial for cw and rapid-scan EPR experiments. The modulation amplitude should not exceed 1/4 of the EPR linewidth to avoid signal distortion (``overmodulation"). Even though in EPR measurements, overmodulation is undesirable, it can be used for precise calibration of a modulation coil. Therefore, an LiPc crystal -- with a linewidth of a few Gauss -- makes it an ideal candidate. The lineshape quickly becomes overmodulated using minimal current in the coil \cite{sojka_sample_2022}. If the overmodulation exceeds the linewidth of the signal, peak-to-peak separation (in Gauss) is equal to the modulation amplitude. By fitting a line to the observed linewidth vs. coil amplitude (in mA), the coil calibration coefficient can be calculated. Calibration coefficients for each sample holder configuration are shown in Table \ref{tab:coil-calibration}.

\begin{figure}[htbp]
\begin{center}
\includegraphics[width=\linewidth]{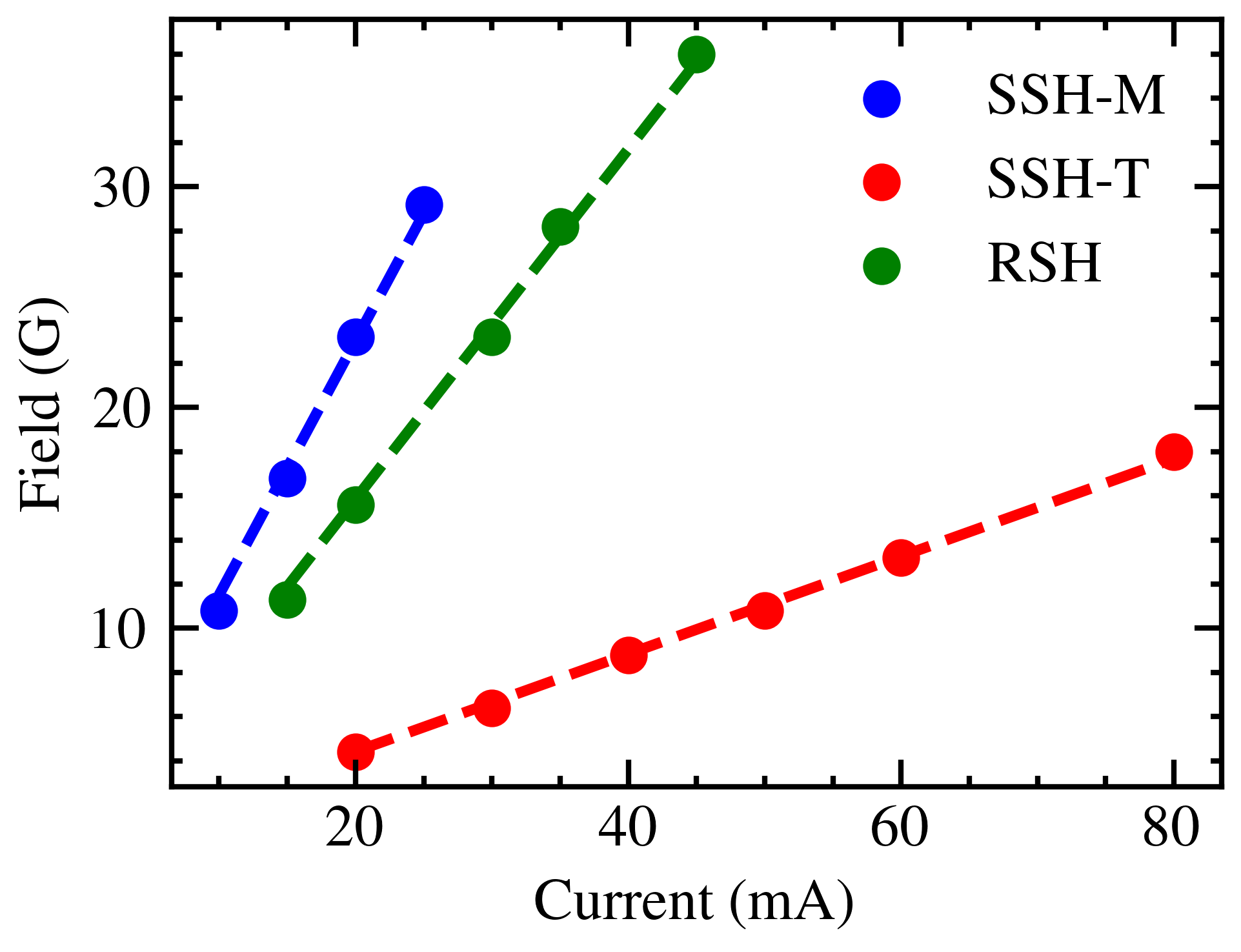}   
\caption{The plot of peak to peak distance of LiPC sample which current used to fed into the modulation coil for each sample holder. The points are fit by linear regression and result in conversion coefficient presented in table \ref{tab:coil-calibration}. }
\label{fig:Modulation}
\end{center}
\end{figure}

It is clear that when the sample was placed at the end of the sample holder, and thus very close to metallic flange of corrugated waveguide, the calibration coefficient was significantly reduced (see Fig. \ref{fig:Modulation}). This phenomena is mainly due to field inhomegeneity caused by placing sample at the edge of the modulation coil as well as magnetic shielding due to eddy currents present in metallic parts of the EPR probe. Therefore, the SSH-T sample position cannot be used for rapid-scan measurements, as sufficient modulation amplitudes cannot be reached at the sample position.

\begin{table}[htbp]
 \caption{Table of calibration coefficients for each sample holder. The coefficients were obtained by fitting a line to the points shown in the Fig. \ref{fig:Modulation}.}
    \centering
   \begin{tabular}{l|c}
        SH type & Calibration coefficient (G/mA)\\
        \hline
        SSH-M &   1.15 \\
        SSH-T &  0.22  \\
        RSH &   0.79 \\
    \end{tabular}
   
    \label{tab:coil-calibration}
\end{table}

\ifCLASSOPTIONcaptionsoff
  \newpage
\fi

\bibliographystyle{IEEEtran}
\bibliography{references_manual}